\documentclass[runningheads]{llncs}
\newcommand{\qedhere}{\qed}

\usepackage[T1]{fontenc}

\usepackage{prettyref}

\newrefformat{sec}{Section\,\ref{#1}}
\newrefformat{app}{Appendix\,\ref{#1}}
\newrefformat{def}{Def.\,\ref{#1}}
\newrefformat{thm}{Theorem\,\ref{#1}}
\newrefformat{prop}{Proposition\,\ref{#1}}
\newrefformat{lem}{Lemma\,\ref{#1}}
\newrefformat{cor}{Corollary\,\ref{#1}}
\newrefformat{rem}{Remark\,\ref{#1}}
\newrefformat{ex}{Example\,\ref{#1}}
\newrefformat{tab}{Table\,\ref{#1}}
\newrefformat{fig}{Fig.\,\ref{#1}}
\newrefformat{case}{case\,\ref{#1}}
\newrefformat{foot}{Footnote\,\ref{#1}}

\usepackage{amsmath}
\usepackage{amssymb}
\usepackage{mathtools}
\usepackage{tensor}
\usepackage{bm}
\usepackage{tikz}
\usepackage{tikz-cd}
\usepackage{graphicx}
\usepackage{stmaryrd}
\usepackage{xspace}
\usepackage{hyperref}

\usepackage{xxcolor}
\usepackage{float}
\usepackage[prefixflatinterpret,prefixflatbracketmodalinterpret,prefixflatsetinterpret,sidenotecalculus,longseqcontext]{logic}
\usepackage[prefixflatinterpret,prefixflatbracketmodalinterpret,differentialdL,simplenames]{dL}

\graphicspath{ {./Images/} }

\definecolor{semblue}{rgb}{0,0,0.7}
\definecolor{vgreen}{rgb}{.1,.5,0}%
\definecolor{vdarkgreen}{rgb}{.06,.3,0}%
\definecolor{vred}{rgb}{.7,0,0}%
\definecolor{vblue}{rgb}{.1,.15,.62}%
\definecolor{vgray}{rgb}{.35,.35,.35}
\definecolor{darkishgray}{rgb}{.35,.35,.35}
\definecolor{vvblue}{rgb}{.14,.21,.868}%
\renewcommand*{\irrulename}[1]{\text{\textcolor{vblue}{\upshape#1}}}%

\newcommand{\R}{\mathbb{R}}
\newcommand{\cR}{\mathcal{R}}
\newcommand{\cU}{\mathcal{U}}
\newcommand{\cV}{\mathcal{V}}

\newcommand{\fR}{\phi_{\cR}}

\newcommand{\Q}{\mathbb{Q}}
\newcommand{\V}{\mathbb{V}}

\newcommand{\X}{\mathcal{X}}

\newcommand{\T}{\mathcal{T}}
\newcommand{\IQ}{\mathbb{I}\mathbb{Q}}
\newcommand{\cA}{\mathcal{A}}

\renewcommand{\P}{\mathbb{P}}

\renewcommand{\S}{\mathbb{S}}

\newcommand{\pv}{\fR \in_{\P} \T}
\newcommand{\mpv}{\models_{\P} \fR \in \T}

\renewcommand{\phi}{\varphi}

\newcommand{\inter}{\text{interior}}

\newcommand{\eps}{\varepsilon}
\newcommand{\eval}[1]{\left\llbracket #1 \right\rrbracket}
\newcommand{\norm}[1]{\left\lVert#1\right\rVert}
\newcommand{\bebecomes}{\mathrel{::=}}
\newcommand{\alternative}{~|~}

\newcommand{\folr}{\textsf{FOL}_{\R}}

\renewcommand*{\der}[1]{\D{(#1)}}

\renewcommand*{\lie}[3][]
{\mathcal{L}_{#2}^{\ifthenelse{\equal{#1}{}}{}{^{\left(#1\right)}}}(#3)}

\renewcommand*{\lied}[3][]{\overset{\bm .}{#3}\ifthenelse{\equal{#1}{}}{}{{}^{(#1)}}}

\renewcommand{\Dostar}[1]{\ifthenelse{\equal{#1}{}}{(*)}{-(*)}}

\newsavebox{\Rval}%
\sbox{\Rval}{$\scriptstyle\mathbb{R}$}
\newsavebox{\backiterateb}%
\sbox{\backiterateb}{$\scriptstyle\overleftarrow{\dibox{{}^*}}$}

\newcommand{\D}[1]{#1'}

\begin{document}

\title{Proof Checking Continuous Reachability}

\author{Long Qian \inst{1} \and Andr\'e Platzer \inst{2}}

\institute{Carnegie Mellon University, Pittsburgh, USA \\
\email{longq@andrew.cmu.edu} \and
Karlsruhe Institute of Technology, Karlsruhe, Germany\\
\email{platzer@kit.edu}}

\maketitle

\begin{abstract}
    This paper introduces a proof format and proof checker for certifying reachability computations of ordinary differential equations (ODEs) with polynomial vector fields.
    Establishing a proof checking framework for certifying ODE reachability is similar to the checking of UNSAT proofs for propositional satisfiability solving, but it is even more important due to the challenges inherent in dealing with ODEs, and even more challenging due to the numerical nature of ODE reachability algorithms. 
    The proofs are expressed in \emph{differential dynamic logic} (\dL), and can be independently verified with symbolic theorem provers for \dL (e.g. \KeYmaeraX). Such proofs are naturally generated by reachability algorithms based on Taylor models, a popular technique to obtain rigorous bounds on approximations of ODEs.
    This paper proves the completeness of such proofs - if an approximation was correctly computed by some (possibly untrusted) reachability algorithm via Taylor models, then a corresponding proof in \dL can be automatically generated and checked to be correct in \dL's proof calculus.
    A dedicated proof checker is implemented in Lean4 for improved performance, which is capable of formally verifying benchmarks in the ARCH competition.
\end{abstract}

\section{Introduction}
\begin{figure}[tbh]
    \centering
    \includegraphics[width=0.75\linewidth]{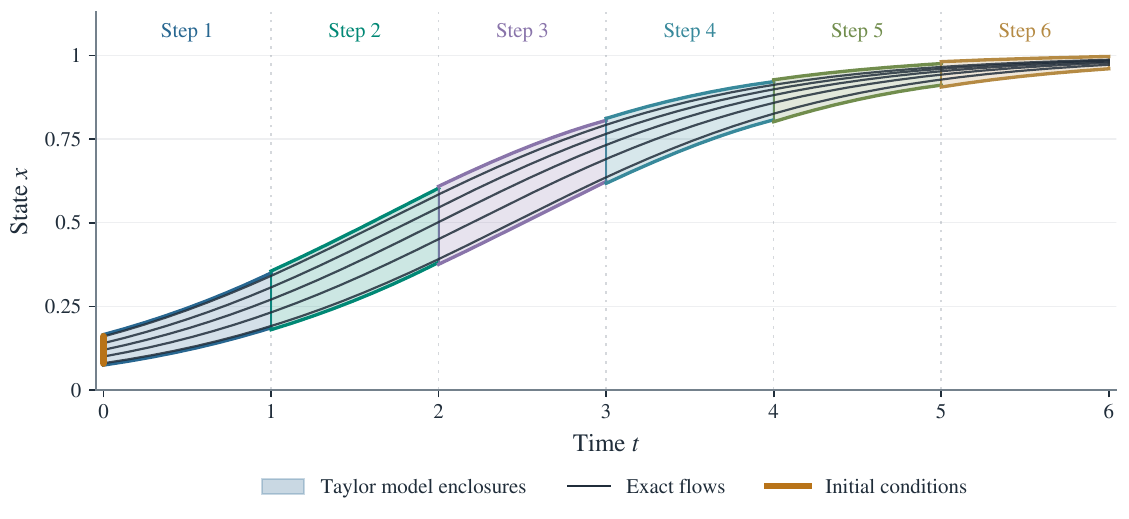}
    \caption{Taylor model overapproximations of an ordinary differential equation over successive time steps, exact trajectories are shown in black.}
    \label{fig:multistep-enclosures}
\end{figure}

This paper concerns the formal verification of reachability computations with approximations for solutions of polynomial ordinary differential equations (ODEs) over compact time horizons (Fig.\,\ref{fig:multistep-enclosures}).
Such problems are of fundamental importance in the verification of Cyber-Physical-Systems (CPS) \cite{Alur_2015,Mitra21} for safety-critical applications, where sound overapproximations can be used to establish safety properties by demonstrating separation between the overapproximations and the unsafe sets. Mathematically, the problem amounts to computing rigorous approximations to the flow function $\phi(x_0, t)$ of time $t$ solving a polynomial differential equation $x' = p(x)$ with initial conditions $x_0 \in \mathcal{X}$ over the compact time horizon $[0, T]$. That is, computing some rigorous approximation $\eta(x_0, t)$ satisfying the following accuracy condition\footnote{In modern implementations such point-wise bounds are generally relaxed to set-valued bounds, detailed in Section\,\ref{sec:proof_checking}.}, where $\eps > 0$ is a desired error bound:
\[\forall x_0 \in \mathcal{X} \, \forall t\in [0, T]~\norm{\phi(x_0, t) - \eta(x_0, t)} < \eps\]

In contrast to standard algorithms for approximating initial value problems (i.e. $\mathcal{X}$ is a singleton set with a concrete numerical vector), reachability problems generally involve uncountably infinite \emph{sets} of initial values, greatly increasing the complexity.
Due to the subtle delicate nature of such algorithms, it is desirable to have machine-checkable guarantees of correctness of the computed approximations for safety-critical applications.
However, such algorithms are very complex and current state-of-the-art implementations \cite{ARCH26:ARCH_COMP26_Category_Report} are designed with performance in mind, across languages such as \verb|C++| \cite{DBLP:conf/adhs/CollinsBGV12,DBLP:conf/cav/ChenAS13}, Matlab \cite{DBLP:conf/cpsweek/Althoff15} and Julia \cite{DBLP:conf/hybrid/BogomolovFFPS19}, which are difficult to formally verify, especially for nontrivial numerical algorithms with floating-point arithmetic \cite{DBLP:journals/corr/abs-2609-19352}.

An interesting parallel can be drawn with the field of SAT solving, where it is infeasible to get state-of-the-art solvers \cite{BiereFallerFazekasFleuryFroleyksPollitt-SAT-Competition-2024-solvers,DBLP:conf/sat/PollittFFFSSB26} both efficient and formally verified.
Instead, such solvers are nowadays required to generate proofs \cite{DBLP:conf/sat/WetzlerHH14}, which can then be checked with formally verified proof checkers \cite{DBLP:journals/sttt/TanHM23} certifying the correctness of the outputs.
Making proof checking happen for rigorous approximations of ODEs poses significant challenges compared to the propositional logic of SAT solving.
It is not even clear in which logic the desired proofs of correctness are best expressed or what a rigorous symbolic proof of a numerical computation for an approximate ODE solution looks like.
Indeed, compared to the obvious choice of propositional logic for SAT solving, one needs to identify a logic of ODEs that is capable of both expressing and certifying the desired proofs of correctness. 

This paper proposes the first proof checking approach to continuous differential equation reachability, applicable to reachability algorithms based on Taylor models \cite{DBLP:conf/snc/MakinoB09}, which are implemented in state-of-the-art tools such as Ariadne \cite{DBLP:conf/adhs/CollinsBGV12}, Flow*  \cite{DBLP:conf/cav/ChenAS13} and JuliaReach \cite{DBLP:conf/hybrid/BogomolovFFPS19}.
Similar to the certification of UNSAT proofs for SAT solvers, such an approach crucially decouples the correctness verification (proof checking) from the computation of the result. The desired approximations of ODEs can be efficiently computed using external implementations, whereas the certification of correctness only requires the result (the approximation with the claimed error bound) and partial information of the computation trace that is already computed by algorithms based on Taylor models. 

More technically, for approximations computed by (potentially unverified) algorithms using the Taylor model architecture, corresponding proofs of correctness can be extracted and expressed in the axiomatic framework of \emph{differential dynamic logic} (\dL) \cite{DBLP:journals/jar/Platzer08,DBLP:journals/jar/Platzer17}, a variant of modal logic designed for the reasoning about hybrid programs.
Furthermore, a completeness result is established, that the axiomatization of \dL \cite{DBLP:conf/lics/Platzer12a,DBLP:journals/jar/Platzer17,DBLP:journals/jacm/PlatzerT20} is complete for proof checking for Taylor models - every correct approximation computed via Taylor models can be deductively proved in \dL. Consequently, this enables the sound implementation of such proof checking procedures in existing theorem provers for \dL (e.g. \KeYmaeraX \cite{Fulton_Mitsch_Quesel_Volp_Platzer_2015}, Isabelle/HOL and Rocq \cite{DBLP:conf/cpp/BohrerRVVP17}, and PVS \cite{dl_pvs}) without affecting the soundness kernel. In summary, the main contributions of this article are:
\begin{enumerate}
    \item A proof format in \dL and proof checking algorithm for certifying approximations of polynomial ODEs computed via Taylor models. 
    \item A completeness theorem for \dL, establishing completeness of the proof format proposed. This result can also be viewed as a contribution to the axiomatic study of ODEs \cite{DBLP:journals/jar/Platzer08,DBLP:conf/lics/Platzer12a,DBLP:journals/jacm/PlatzerT20,DBLP:journals/jacm/PlatzerQ25}, which is of independent interest. 
    \item An implementation in \KeYmaeraX to check such proofs, without modifications to the soundness-critical kernel. 
    \item An implementation in Lean4 that is specialized to the certification of such proofs, attaining efficient performance. This implementation is capable of certifying safety properties from the most recent ARCH competition \cite{ARCH26:ARCH_COMP26_Category_Report}. 
\end{enumerate}

As a concrete demonstration of the impact of using proof checkers, we uncovered soundness bugs in two implementations of Taylor model based reachability algorithms, highlighting the intricacy of such procedures and ODE reasoning.

\section{Background: Differential Dynamic Logic}
\emph{Differential dynamic logic} (\dL) is a variant of the classical modal logic designed to reason about hybrid programs, supporting the deductive verification of CPS. 
\subsection{Syntax}
The continuous fragment of \dL needed in this paper admits the following syntax of terms, formulas, and continuous programs (where $c \in \Q$ and ${\preceq} \in \{{<},{\leq}\}$):
\begin{align*}
    p, q &\bebecomes x\alternative c\alternative p + q\alternative p\cdot q\\
    \phi, \psi &\bebecomes p \preceq q \alternative \phi \land \psi \alternative \phi \lor \psi \alternative \neg \phi \alternative \forall x \phi \alternative \exists x \phi \alternative 
\dbox{\alpha}{\psi}\\
\alpha &\bebecomes \cdots \alternative x' = p(x) \,\&\, Q
\end{align*}
Intuitively, the terms of \dL are multivariate polynomials with rational coefficients, the formulas consist of standard constructions with the usual first-order logic of real arithmetic $\folr$ interpretation, and the additional (safety) modal formula $\dbox{\alpha}{\phi}$ where $\alpha \equiv x' = p(x) \,\&\, Q$ is some ODE system with a \emph{domain constraint} $Q$ in $\folr$. In this paper, $x = (x_0, \dots, x_n)$ denotes a vectorial variable. 

\subsection{Semantics and Axiomatization}
The semantics of \dL are denotational. A state $\omega$ is a mapping $\omega : \V \to \R$ that assigns a real value to every variable, where $\V$ denotes the set of variables, and $\S$ denotes the set of all such states. Terms are evaluated as polynomials at the state $\omega$, and the semantics of first-order logical connectives and quantifiers are defined as usual. For a \dL formula $\phi$, its semantics $\eval{\phi}$ is the set of all states $\omega \in \S$ such that $\omega \models \phi$, i.e. the formula $\phi$ is true in $\omega$.

For an ODE $\alpha \equiv x' = p(x) \,\&\, Q$ and a state $\omega \in \S$, let $\Phi_\omega : [0, T_\omega) \to \S$ be the solution to $x' = p(x)$ extended maximally to the right, with $0 < T_\omega \leq \infty$ and $\Phi_\omega(0) = \omega$. Here $x$ is understood to be vectorial, and variables not bound by the ODE remain unchanged. The formula $\dbox{\alpha}{\phi}$ is true in $\omega$ iff for every $0 \leq \tau < T_\omega$ such that $\Phi_\omega(s) \models Q$ for all $s \in [0, \tau]$, we have $\Phi_\omega(\tau) \models \phi$.

Intuitively, $\dbox{\alpha}{\phi}$ expresses a \emph{safety} property, that $\phi$ holds after every evolution of the ODE staying within the domain constraint $Q$. Evolutions of duration zero are included whenever $Q$ holds initially. Finally, a formula $\phi$ is said to be valid, written $\models \phi$, if $\eval{\phi} = \S$, i.e. it is true in all states. 
\begin{theorem}[\cite{DBLP:conf/lics/Platzer12a,DBLP:journals/fac/TanP21,DBLP:journals/jacm/PlatzerQ25}]
    \label{thm:dL_proof_rules}
    The following (derived) proof rules of \dL are sound.

    \begingroup
    \renewcommand{\linferenceRuleNameSeparation}{\hspace{1em}}
    \setlength{\linferenceRulevskipamount}{1.4em}
    \begin{center}
    \begin{calculus}
        \cinferenceRule[dI|dI]{differential invariant}
        {\linferenceRule[sequent]
            {\lsequent[g]{\ivr}{\der{F}_{\D{x}=\genDE{x}}}}
            {\lsequent{F}{\dbox{\pevolvein{\D{x}=\genDE{x}}{\ivr}}{F}}}
        }{}

        \cinferenceRule[dC|dC]{differential cut}
        {\linferenceRule[sequent]
            {\lsequent[L]{}{\dbox{\pevolvein{\D{x}=\genDE{x}}{\ivr}}{\cusfml}}
            &\lsequent[L]{}{\dbox{\pevolvein{\D{x}=\genDE{x}}{(\ivr\land \cusfml)}}{\ousfml[x]}}}
            {\lsequent[L]{}{\dbox{\pevolvein{\D{x}=\genDE{x}}{\ivr}}{\ousfml[x]}}}
        }{}

        \cinferenceRule[enclosure|Enc]{topological enclosure}
        {\linferenceRule[sequent]
            {\lsequent{\Gamma}{e \geq 0}
            &\lsequent{\Gamma}{\dbox{x' = f(x)\,\&\,Q \land e \geq 0}{e > 0}}}
            {\lsequent{\Gamma}{\dbox{x' = f(x)\,\&\,Q}{e > 0}}}
        }{}

    \end{calculus}
    \end{center}
    \endgroup
\end{theorem}

Intuitively, the differential invariant rule \irref{dI} allows one to prove that a formula $F$ is an invariant of an ODE by proving that its Lie derivative $\der{F}$ is satisfied in the domain constraint $\ivr$ after substituting in $x' \mapsto f(x)$. The differential cut rule \irref{dC} allows one to strengthen the domain constraint of an ODE by \emph{cutting in} some formula $\cusfml$ which is an invariant. The topological enclosure rule \irref{enclosure} allows one to prove that a formula $e > 0$ holds along an ODE by proving that it holds initially and that it is preserved along the ODE in the domain constraint strengthened with $e \geq 0$. The justification of \irref{enclosure} is surprisingly subtle, and implicitly uses the completeness of the real numbers \cite{DBLP:journals/jacm/PlatzerT20,DBLP:journals/fac/TanP21}. 

\section{Proof Checking for Continuous Reachability}
\label{sec:proof_checking}

This section develops a proof checking framework in \dL for continuous reachability problems along differential equations based on Taylor models, as well as establishing corresponding completeness results. The following definition gives the class of reachability problems over compact time horizons.

\begin{definition}[Continuous Reachability Problems]
    \label{def:reachability_problem}
    A \emph{reachability problem} is a triple $\cR = (\X, p, T)$ where:
    \begin{itemize}
        \item $\X \subset \R^n$ is a compact set of initial conditions definable in $\folr$.
        \item $p \in \Q[x]^n$ is a polynomial vector field with rational coefficients.
        \item $T \in \Q^+$ represents the time horizon $[0, T]$.
    \end{itemize}
    The \emph{flow} of $\cR$, denoted $\fR : \X \times [0, T] \to \R^n$, is the unique function\footnote{In general $\phi_{\cR}$ might not be well-defined for all $T>0$ due to finite-time blow up. However, in this paper $\phi_{\cR}$ is only used when a Taylor model enclosure has been established, hence bounded.} such that
    \[\phi_{\cR}'(x, t) = p(\phi_{\cR}(x, t)) \qquad \phi_{\cR}(x, 0) = x\]
\end{definition}
Intuitively, a reachability problem $\cR = (\X, p, T)$ corresponds to the problem of computing rigorous approximations of the flow $\fR$. In \dL's syntax, formally verifying continuous reachability approximations amounts to deductively proving formulas of the following form, where $S(x)$ is some desired safety condition, e.g. accuracy of some approximation, or avoidance of a pre-specified unsafe region.
\[x \in \X \land t = 0 \rightarrow \dbox{x' = p(x), t' = 1 \& t \leq T}{S(x)}\]
This paper is primarily concerned with certifying reachability algorithms based on Taylor models \cite{DBLP:conf/snc/MakinoB09}. That is, suppose that some rigorous enclosure of the flow $\fR$ has been obtained using Taylor models computed via some external implementation, how can one verify the correctness of this enclosure in \dL by proving the formula above. 

The following definition introduces the primitive object of a Taylor model. In a nutshell, a Taylor model is a multivariate polynomial with a remainder interval, approximating the flow of an ODE in some reachability problem. 

\begin{definition}[Taylor Model Enclosures]
    \label{def:valid_enclosure}
    A \emph{Taylor model} is a pair $\T = (r, I)$ where:
    \begin{itemize}
        \item $r \in \Q[x]^n$ is a (vectorial) multivariate polynomial with rational coefficients.
        \item $I \in \IQ^n$ is a (vectorial) rational remainder interval.
    \end{itemize}
    $\T$ is said to be a \emph{valid} approximation of the reachability problem $\cR = (\X, p, T)$, denoted $\fR \in \T$, if the following holds.
    \[\forall x_0\!\in\!\X\, \forall t\!\in\![0, T]~\fR(x_0, t) - r(x_0, t) \in I\]
\end{definition}

A Taylor model $\T = (r, I)$ is also identified with the natural set-valued map $x \mapsto r(x) + I$. Since Taylor models and reachability problems are all \dL-definable, the validity condition $\fR \in \T$ can be naturally expressed in \dL as well. 

\begin{theorem}[Sound Enclosure in \dL]
    Given a Taylor model $\T = (r, I)$ and a reachability problem $\cR = (\X, p, T)$, the enclosure condition $\fR \in \T$ is equivalent to the validity of the following \dL formula 
    \[x_0 \in \X \land x = x_0 \land t = 0 \rightarrow \dbox{x' = p(x), t' = 1 \& t \leq T}{x - r(x_0, t) \in I}\]
\end{theorem}
Consequently, the (semantic) condition $\fR \in \T$ is equivalently identified with its corresponding \dL formula, and sometimes also written as $\models \fR \in \T$ to emphasize its validity as a \dL formula.

In a nutshell, the core behind reachability algorithms based on Taylor models is that they decompose the desired time interval $[0, T]$ into time steps $[0=t_0, t_1]$, $\dots, [t_{N - 1}, t_{N} = T]$ and synthesize a Taylor model on each interval representing some rigorous enclosure, while checking correctness of the remainder intervals by checking contraction of the Picard operator \cite{DBLP:conf/snc/MakinoB09}. State-of-the-art algorithms \cite{DBLP:conf/cav/ChenAS13,DBLP:conf/adhs/CollinsBGV12,DBLP:conf/hybrid/BogomolovFFPS19} further adopt sophisticated optimizations such as pre-conditioning/reconditioning \cite{tm_neher_jackson_nedialkov,DBLP:conf/adhs/CollinsBGV12} for improved performance. 

\begin{figure}[tb]
  \centering
  \includegraphics[width=\linewidth]{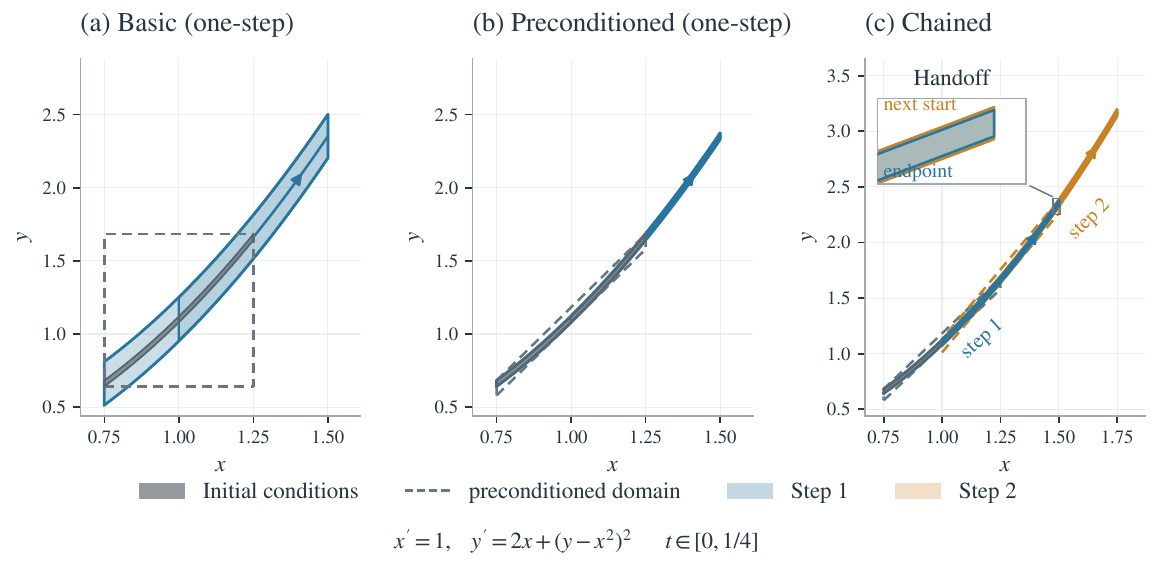}
  \caption{Basic one-step, preconditioned one-step, and chained Taylor
  model enclosures for $x'=1$ and $y'=2x+(y-x^2)^2$.
  Both one-step enclosures use the same initial conditions.}
  \label{fig:nonlinear-preconditioning}
\end{figure}

The rest of the section proceeds as follows: Section\,\ref{sec:one_step_enclosure} establishes completeness results of Taylor model enclosures over one time step. Theorem \ref{thm:one_step_completeness} establishes completeness for basic Taylor model enclosures (Fig.\ref{fig:nonlinear-preconditioning}, panel 1). Theorem \ref{thm:preconditioned_completeness} establishes completeness for preconditioned Taylor model enclosures (Fig.\ref{fig:nonlinear-preconditioning}, panel 2), these are necessary to obtain tight enclosures and are implemented in practical tools such as Flow* \cite{DBLP:conf/cav/ChenAS13} and Ariadne \cite{DBLP:conf/adhs/CollinsBGV12}. Section\,\ref{sec:complete_enclosures} establishes completeness results for a sequence of (preconditioned) Taylor model enclosures over multiple time steps (Fig.\ref{fig:nonlinear-preconditioning}, panel 3), providing a complete proof checking framework for continuous reachability problems based on Taylor models.

\subsection{One-Step Enclosures}
\label{sec:one_step_enclosure}

Mathematically, the use of Taylor models to obtain rigorous enclosures of ODEs \cite{DBLP:conf/snc/MakinoB09,tm_neher_jackson_nedialkov} amounts to checking a contractive condition of the Picard operator, commonly stated for reachability problems with a single initial value as follows. 

\begin{definition}
    Given a (singleton) reachability problem $\cR = (\{x_0\}, p, T)$, the \emph{Picard operator} $\P_{x_0} : C^0([0, T], \R^n) \to C^0([0, T], \R^n)$ is defined by 
    \[(\P_{x_0}f)(t) = x_0 + \int_{0}^t p(f(s)) ds\]
    It is a well-known result of the Picard-Lindel\"of theorem that the solution $s(t)$ of the ODE\footnote{Assuming it exists over $[0, T]$} is the unique fixed point $\P_{x_0}(s) = s$. 
\end{definition}

However, in practical applications such singleton reachability problems are insufficient, since already after one step of an approximation, the resulting overapproximation is a \emph{set} of possible values. Hence, the Picard operator needs to be generalized to act on sets of initial values. Such generalizations are also necessary for the analysis of CPS, where initial values are generally set-valued. 

\begin{definition}[Picard Image \cite{DBLP:conf/snc/MakinoB09}]
    Let $\cR = (\X,p,T)$ be a reachability problem and
    $\T=(r,I)$ a Taylor model. For each fixed input $z \in \R^n$, define
    \[
        \T_z \coloneqq
        \left\{
            f \in C^0([0,T],\R^n)
            \;\middle|\;
            \forall t\in[0,T]\;
            f(t)-r(z,t)\in I
        \right\}.
    \]
    The \emph{Picard image} of $\T$ relative to $\cR$ is
    \[
        \P_{\cR}(\T)
        \coloneqq
        \bigcup_{z\in\X}
        \left\{(z,\P_z(f)) \mid f\in\T_z\right\}.
    \]
\end{definition}

That is, the Picard image of $\T$ relative to $\cR$ is the union of the pointwise images over $\X$ labeled with corresponding initial values. Importantly, it then follows from Schauder's fixed point theorem \cite{Walter_1998} that $\T$ soundly contains the true flow when it is contractive under the Picard operator, establishing $\fR \in \T$.

\begin{theorem}[\cite{DBLP:conf/snc/MakinoB09}]
    \label{thm:contractive_picard}
    Let $\cR=(\X,p,T)$ be a reachability problem and
    $\T=(r,I)$ a Taylor model. Suppose that
    \[
        \forall (z,f)\in\P_{\cR}(\T)\;
        \forall t\in[0,T]\;
        f(t)-r(z,t)\in I.
    \]
    Then $\fR\in\T$. This condition is abbreviated as
    $\P_{\cR}(\T)\subseteq\T$, and the theorem can be succinctly stated as:
    \[
        \P_{\cR}(\T) \subseteq \T \implies \fR \in \T
    \]
\end{theorem}
Intuitively, the hypothesis of Theorem \ref{thm:contractive_picard} states that for all $z \in \X$, the image of the Picard operator $\P_z$ on the set of functions $\T_z$ is again contained in $\T_z$. I.e. $\P_z(\T_z) \subseteq \T_z$ (hence the abbreviation), which allows one to apply Schauder's fixed point theorem to bound $\fR$ \cite{DBLP:conf/snc/MakinoB09}. The following example demonstrates how Theorem \ref{thm:contractive_picard} can be used to establish a Taylor model enclosure of a reachability problem.

\begin{example}[One-step Enclosure]
    \label{ex:basic-picard-enclosure}
    \begin{figure}[tbh]
    \centering
    \includegraphics[width=0.75\linewidth]{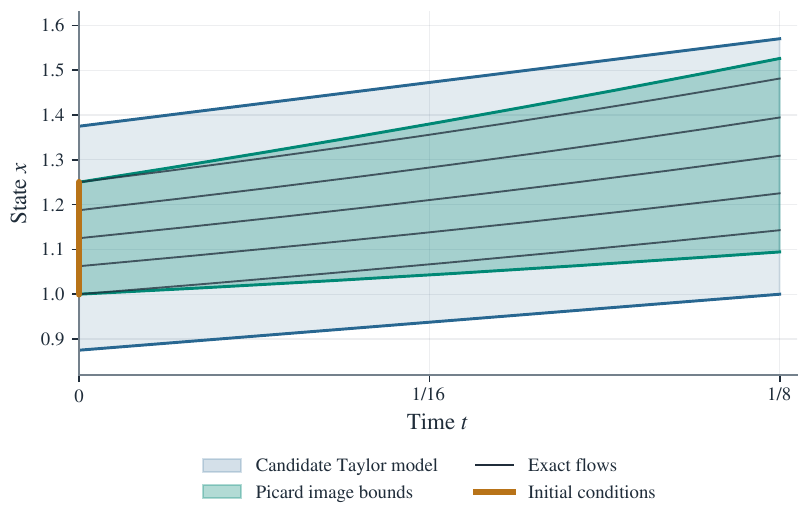}
    \caption{One-step enclosure for $\cR$.
        Candidate Taylor model (blue) strictly contains
        the computed bounds on its Picard image (green).}
        \label{fig:basic-picard-enclosure}
    \end{figure}
    Let $\cR=(\X,p,T)$ be a reachability problem with
    $p(x)=x^2$, initial conditions $\X=[1,5/4]$, and $T=1/8$.
    Choose the Taylor model
    \[
        \T=(r,I)\qquad
        r(z,t)=z+z^2t,\qquad I=[-1/8,1/8]
    \]
    Fix $z\in\X$ and $f\in\T_z$, where
    $f(t)=r(z,t)+e(t)$ with $e(t)\in I$.
    On $\X\times[0,T]$ numerical bounding gives $1\leq r(z,t)\leq3/2$,
    interval arithmetic gives
    \[
        2r(z,t)e(t)\in[-3/8,3/8]
        \qquad e(t)^2\in[0,1/64]
    \]
    Hence the vector-field image is bounded by another
    Taylor model $(q,J)$:
    \[
        p(f(t))\in q(z,t)+J,\qquad
        q(z,t)=r(z,t)^2,\qquad
        J=[-3/8,25/64]
    \]
    Directly integrating these bounds gives
    \begin{align*}
        (\P_z f)(t) &\in z+\int_0^t q(z,s)\,ds+tJ = r(z,t)+z^3t^2+\frac{z^4t^3}{3}+tJ
    \end{align*}
    Direct bounds with interval arithmetic then give
    \[
        (\P_z f)(t)-r(z,t)
        \in[-3/64,21/256]
        \subset[-1/8,1/8]
    \]
    By Theorem~\ref{thm:contractive_picard}, this establishes validity of $\fR \in \T$. 
\end{example}

To the best of our knowledge, most implementations of Taylor model enclosures rely on some version of Theorem \ref{thm:contractive_picard} to ensure validity. Utilizing conservative bounding to check for contraction by first bounding the vector field image $p(\T)$ with another Taylor model, and then evaluating the resulting bounds as in Example \ref{ex:basic-picard-enclosure}. As such, this paper focuses on these Picard-valid Taylor models with a slight strengthening, recorded in the following definition.

\begin{definition}[Picard-Valid Taylor Model]
    \label{def:picard_valid_basic_model}
    A Taylor model $\T$ is said to be \emph{Picard-valid} for a reachability problem $\cR$ if there exists some
    $\eps \in \Q^+$ such that, for every
    $(z,f)\in\P_{\cR}(\T)$ and $t\in[0,T]$,
    \[
        f(t)-r(z,t)+[-\eps,\eps]^n
        \subseteq I.
    \]
    Equivalently, the $\eps$-enlarged Picard image containment $\P_{z}(\T_z) + [-\eps,\eps]^n \subseteq \T_z$ holds at every $z \in \X$. This condition is denoted by $\pv$, or $\mpv$ when emphasizing logical validity.
\end{definition}

While Definition \ref{def:picard_valid_basic_model} is a slight strengthening of condition $\P_{\cR}(\T) \subseteq \T$, it does not affect practical applicability.  Since in practice Theorem \ref{thm:contractive_picard} is always checked using conservative bounding with floating-point arithmetic, and hence some $\eps$-enlargement is always present. On the other hand, the $\eps$-enlargement allows for a more uniform presentation of the following completeness theorem. Establishing that any Picard-valid Taylor model is actually provably valid in \dL. The key ideas are sketched, with the full proof provided in Appendix \ref{app:proofs}.

\begin{theorem}[One-Step Completeness]
    \label{thm:one_step_completeness}
    Let $\cR$ be a reachability problem and $\T$ a Picard-valid Taylor model for $\cR$, then $\fR \in \T$ is provably valid using the standard axiomatization of \dL:
    \[\mpv ~\implies~ \vdash_{\dL} \fR \in \T\]
\end{theorem}
\begin{proof}[Sketch]
Recall that the goal is to derive the following \dL formula from the Picard-validity assumption $\mpv$ (recall $\cR = (\X, p, T), \T = (r, I)$):
\[x = x_0 \land x_0 \in \X \land t = 0 \rightarrow \dbox{x' = p(x), t' = 1 \& t \leq T}{x - r(x_0, t) \in I}\]
The main strategy of the proof is the following :
\begin{itemize}
    \item Utilizing the $\eps$-containment assumption of Picard-validity (Definition \ref{def:picard_valid_basic_model}), one can establish an overapproximation\footnote{Due to technical reasons the interval $I'$ might have polynomial endpoints in $x_0, t$, hence not strictly a Taylor model. However, the rest of the proof is not affected by this.} $\T' = (r', I')$ of $p(\T)$ (as set-valued maps) such that the following condition holds for all $z \in \X, t\in [0, T]$. 
        \[S(z,t) - r(z, t) + \hat{I} \subseteq \inter(I)\]
    where $\hat{I}$ is the time-integral of $I'$ and
    \[ S(z,t) = z + \int_0^t r'(z,s)\,ds\]
    \item Since integrals of polynomials are again polynomials, the condition above can be expressed as a $\folr$ formula in $\dL$ (dependent on $\T'$).
    \item Crucially, the proof proceeds by cutting in the desired constraint with the proof rule \irref{enclosure}. 
        \[x - r(x_0, t) \in I\]
    \item The remaining non-trivial obligation arising from \irref{enclosure}
    is to prove $x-r(x_0,t)\in\inter(I)$ under the evolution
    domain strengthened with $x-r(x_0,t)\in I$.
    \item 
    By construction of $\T'$ and the domain constraint $x \in \T(x_0, t)$, this gives $p(x)-r'(x_0,t)\in I'$. Hence \irref{dI} establishes the invariant
    \[
        x-S(x_0,t)\in \hat{I}
    \]
    The desired proof can then be completed by cutting this in with \irref{dC}.
    \qedhere
\end{itemize}
\end{proof}
Theorem \ref{thm:one_step_completeness} is somewhat surprising, since the axiomatization of \dL does not include fixed point properties such as Picard/Schauder's fixed point theorem. However, by clever usage of the proof rules \irref{dI+enclosure}, \dL can establish a deductive proof for any Picard-valid enclosure. Indeed, the key takeaway is that the essence of Schauder's fixed point theorem (in relation to Theorem \ref{thm:contractive_picard}) is captured by the topological enclosure rule \irref{enclosure}, which follows from the completeness of $\R$.

While Theorem \ref{thm:one_step_completeness} is in theory sufficient to establish the desired proofs in \dL certifying one-step enclosures, more sophisticated implementations \cite{DBLP:conf/cav/ChenAS13,DBLP:conf/adhs/CollinsBGV12} of Taylor models generally use a pair of preconditioned Taylor models in a factorized form to obtain tighter enclosures (Fig.\ref{fig:nonlinear-preconditioning}), given in the following definition. 

\begin{definition}[Preconditioned Taylor Model \cite{tm_neher_jackson_nedialkov}]
    A \emph{preconditioned Taylor model} $\T = \cU \circ \cV$ is the \emph{composition} of two Taylor models, the \emph{left Taylor model}, $\cU = (r_{\cU}, I_{\cU})$, and the \emph{right Taylor model}, $\cV = (r_{\cV}, I_{\cV})$. $\T$ is also identified as its corresponding set-valued map
    \[\T(x, t) \coloneqq \cU(\cV(x), t)\]
\end{definition}

Intuitively, preconditioned Taylor models represent a change-of-basis in the spatial coordinates, allowing for better approximations when integrating through the local flow $\cU$. The right Taylor model $\cV$ represents the transformation from the original coordinates to the preconditioned coordinates, and the left Taylor model $\cU$ represents the local flow in the preconditioned coordinates. Importantly, preconditioned Taylor models can be relaxed to give \emph{set-valued} approximations. The following definition makes precise the validity of preconditioned Taylor models. 

\begin{definition}[Valid Preconditioned Taylor Model] \label{def:valid-preconditioned-Taylor}
    Let $\cR = (\X, p, T)$ be a reachability problem and $\T = \cU \circ \cV$ be a preconditioned Taylor model. $\T$ is \emph{valid} if there exists a $\folr$ definable set $B$ such that
    \[\forall t \in [0, T]~\fR(\X, t) \subseteq \T(B, t) = \cU(\cV(B), t)\]
    Equivalently, $\T$ is \emph{valid} if and only if the following \dL formula is valid
    \[x \in \X \land t = 0 \rightarrow \dbox{x' = p(x), t' = 1 \& t \leq T}{x \in \T(B, t) = \cU(\cV(B), t)}\]
    In either (hence both) case, validity is also denoted as $\fR \in \T$, or $\models \fR \in \T$, similar to Definition \ref{def:picard_valid_basic_model}. 
\end{definition}
Thus, a preconditioned Taylor model $\T = \cU \circ \cV$ is valid if there exists some domain $B$ such that the set-valued map $t \mapsto \cU(\cV(B), t)$ is a sound overapproximation of the set-valued flow $t \mapsto \fR(\X, t)$ over $[0, T]$. A validity theorem similar to Theorem \ref{thm:contractive_picard} can be established for preconditioned Taylor models.

\begin{theorem}[Picard-Valid Preconditioned Taylor Model \cite{tm_neher_jackson_nedialkov}]
    \label{thm:preconditioned_validity}
    Let $\cR = (\X, p, T)$ be a reachability problem and let $\T = \cU \circ \cV$ be a preconditioned Taylor model, with $\cU = (r_{\cU}, I_{\cU})$ and $\cV = (r_{\cV}, I_{\cV})$. Suppose that there exist (topologically compact) $\folr$ definable sets $B, D$ and a (possibly trivial) Taylor model $\cA$, denoted the scaling map, such that the following conditions hold:
    \begin{itemize}
        \item Preconditioning:
        \[\X \subseteq \cA(\cV(B)) \qquad\qquad \cV(B) \subseteq D\]
        \item Local Picard-validity: For all $z \in D$, one has
            \[\P_{\cA(z)}(\cU_z) + [-\eps,\eps]^n = \bigcup_{x_0 \in \cA(z)} \P_{x_0}(\cU_z)  + [-\eps,\eps]^n\subseteq \cU_z\]
            for some uniform $\eps \in \Q^+$. 
    \end{itemize}
    Then the enclosure conclusion $\fR \in \T = \cU \circ \cV$ holds, and $\T$ is said to be \emph{Picard-valid}, denoted $\pv$ or $\mpv$, similar to Definition \ref{def:picard_valid_basic_model}. 
\end{theorem}

Similar to Theorem \ref{thm:completeness_of_tm}, a completeness result can be established for Picard-valid preconditioned Taylor models. The proof is similar to the proof of Theorem \ref{thm:one_step_completeness}, and is provided in Appendix \ref{app:proofs}.

\begin{theorem}[Completeness for Preconditioned Taylor Models]
    \label{thm:preconditioned_completeness}
    Let $\cR$ be a reachability problem and $\T = \cU \circ \cV$ be a Picard-valid preconditioned (Theorem \ref{thm:preconditioned_completeness}) Taylor model for $\cR$, then $\fR \in \T$ is provably valid using the standard axiomatization of \dL:
    \[\mpv ~\implies~ \vdash_{\dL} \fR \in \T = \cU \circ \cV\]
\end{theorem}

Consequently, the proof format for Taylor model enclosures, localized to each step $[t_{i}, t_{i+1}]$, will be a preconditioned Taylor model, accompanied by the witnesses of Picard-validity in Theorem \ref{thm:preconditioned_validity}. 

\subsection{Complete Enclosures}
\label{sec:complete_enclosures}

Section\,\ref{sec:one_step_enclosure} established the desired proof format and completeness theorem for Taylor model enclosures on one time step $[t_i, t_{i+1}]$. This section completes the argument by chaining up the proof data across all local steps, by provably certifying a trace of one-step Taylor models (Figures \ref{fig:multistep-enclosures} and \ref{fig:nonlinear-preconditioning}).

\begin{definition}[Taylor Model Trace]
    \label{def: Taylor Model Trace}
    Let $\cR = (\X, p, T)$ be a reachability problem. A \emph{Taylor model trace} consists of:
    \begin{itemize}
        \item A sequence of preconditioned Taylor models $(\T_0, \dots, \T_N)$. 
        \item A sequence of time steps $[0 = t_0, t_1], \dots, [t_{N}, t_{N+1} = T]$.
    \end{itemize}
\end{definition}

Intuitively, a Taylor model sequence represents a sequence of enclosures by the Taylor models $\T_i$ over the time steps $[t_i, t_{i + 1}]$. Generalizing the one-step case gives the following notion of validity.

\begin{definition}[Valid Taylor Model Trace]
    \label{def:valid_trace}
    Let $\cR = (\X, p, T)$ be a reachability problem, and let $\T = (\T_0, \dots, \T_N)$ be a Taylor model trace. $\T$ is said to be \emph{valid} if there exist $\folr$ definable sets $(B_0, \dots, B_N)$ such that
    \[\forall t \in [0, T]~\bigwedge_{i} \left(t \in [t_{i}, t_{i + 1}] \rightarrow \fR(\X, t) \subseteq \T_i(B_i, t - t_i)\right)\]
    That is, the trace is valid if each step of the trace is a valid preconditioned enclosure in the sense of Definition \ref{def:valid-preconditioned-Taylor}.
     Equivalently, $\T$ is valid if and only if the following \dL formula is valid
    \[x \in \X \land t = 0 \rightarrow \dbox{x' = p(x), t' = 1 \& t\leq T}{\bigwedge_i \left(t \in [t_{i}, t_{i + 1}] \rightarrow x \in \T_i(B_i, t - t_i)\right)}\]
    In either (hence both) case, validity is also denoted as $\fR \in \T$, or $\models \fR \in \T$
\end{definition}

Intuitively, a Taylor model trace is valid if each individual step is valid on its corresponding step, thereby establishing a sound overapproximation:
\[\fR(\X, [0, T]) \subseteq \bigcup_{i} \T_i(B_i, [0, t_{i + 1} - t_i])\]
As in the one-step case, the following theorem relates validity of the entire trace to the Picard-validity of the individual steps. 

\begin{theorem}[Picard-Valid Taylor Model Trace]
    \label{thm:picard_valid_trace}
    Let $\cR$ be a reachability problem and $(\T_0, \dots, \T_N)$ a Taylor model trace with time steps $[0 = t_0, t_1], \dots, [t_{N}, t_{N+1} = T]$. Further suppose that there exists a sequence of definable sets $(E_0, \dots, E_N)$ such that: 
    \begin{itemize}
        \item Picard-validity holds at each step:
            \[\bigwedge_{i} \phi_{\cR_i} \in_{\P} \T_i\]
    where $\cR_i = (E_i, p, t_{i + 1} - t_i)$ is the reachability problem for the $i$-th step. 
        \item Sound containment of endpoints holds:
            \[\bigwedge_{i > 0} \T_{i - 1}(B_{i - 1}, t_{i} - t_{i - 1}) \subseteq E_{i}\]
            where $\T_i = \cU_i \circ \cV_i$ is the $i$-th valid preconditioned Taylor model for $\cR_i$, with domain $B_i$. And the initial condition $\X \subseteq E_0$. 
    \end{itemize}
    Then the enclosure conclusion $\fR \in \T = (\T_0, \dots, \T_N)$ holds, and $\T$ is said to be \emph{Picard-valid}, denoted $\pv$ or $\mpv$.
\end{theorem}

Theorem \ref{thm:picard_valid_trace} gives a natural definition of proofs for reachability enclosures computed via Taylor models. To the best of our knowledge, the proposed certificate format encapsulates standard implementations in practical tools, such as Ariadne \cite{DBLP:conf/adhs/CollinsBGV12}, Flow* \cite{DBLP:conf/cav/ChenAS13} and JuliaReach \cite{DBLP:conf/hybrid/BogomolovFFPS19}. Similar to the one-step case, the following completeness theorem establishes that Picard-valid Taylor model traces are always \dL provable, the proof is provided in Appendix \ref{app:proofs}.

\begin{theorem}
    \label{thm:completeness_of_tm}
    Let $\cR = (\X, p, T)$ be a reachability problem, and let $\T = (\T_0, \dots, \T_N)$ be a Picard-valid Taylor model trace. Then the corresponding \dL formula in Theorem \ref{thm:picard_valid_trace} is provable:
    \[\models_{\P} \fR \in (\T_0, \dots, \T_N) ~\implies~ \vdash_{\dL} \fR \in (\T_0, \dots, \T_N)\]
\end{theorem}

Finally, Theorem \ref{thm:completeness_of_tm} demonstrates that given data witnessing Picard-validity, one can always generate a corresponding proof in \dL, formally certifying the correctness of the approximation (assuming Picard-validity). For the sake of precision, the following definition gives a formal notion of a Picard-certificate for a Taylor model trace.

\begin{definition}[Picard-Certificate]
    \label{def:picard_certificate}
    Let $\cR = (\X, p, T)$ be a reachability problem, and let $\T = (\T_0, \dots, \T_N)$ be a Picard-valid Taylor model trace over time steps $(t_0, \dots, t_N)$. A \emph{Picard-certificate} for $\T$ is:
    \begin{itemize}
        \item A sequence of definable sets $(E_i)_i$ witnessing the Picard-validity of $\T$ in the sense of Theorem \ref{thm:picard_valid_trace}.
        \item A sequence of definable sets $(B_i, D_i)_{i}$, where $B_i$ is the domain of the $i$-th right Taylor model $\cV_i$, and $D_i$ is the domain of the $i$-th left Taylor model $\cU_i$, together with a sequence of Taylor models (scaling maps) $(\cA_i)_i$ witnessing the Picard-validity of $\T_i$ in the sense of Theorem \ref{thm:preconditioned_validity}.
    \end{itemize}
\end{definition}
\begin{example}[2D Van der Pol]
The first step for the 2D Van der Pol system $x' = y, y' = (1 - x^2)y - x$ (Appendix \ref{app:additional_experimental_results}, Table \ref{tab:benchmark-goals}) has the following
Picard-certificate computed by Ariadne, with coefficients rounded and polynomial
terms of time degree at least three abbreviated by $O(t^3)$.
{\small
\[
    E_0=[1.25,1.55]\times[2.35,2.45]
    \qquad B_0=D_0=[-1,1]^2 \qquad [t_0,t_1]=[0,h]\qquad h\approx0.02
\]
}
with preconditioned Taylor model 
{
\small
\begin{align*}
    \cA_0(z) \approx&
    \begin{pmatrix}1.4+0.15z_1\\2.4+0.05z_2\end{pmatrix}\\
    \cV_0(b)\approx&
            \begin{pmatrix}
                (1-3.179\cdot10^{-7})b_1\\
                (1-1.907\cdot10^{-6})b_2
            \end{pmatrix} +[-3.179\cdot10^{-7},3.179\cdot10^{-7}]
            \times[-1.907\cdot10^{-6},1.907\cdot10^{-6}]\\
    \cU_0(z,t)\approx &
        \begin{pmatrix}u_1(z,t)\\u_2(z,t)\end{pmatrix}+[-4.747\cdot10^{-9},4.747\cdot10^{-9}]
        \times[-9.908\cdot10^{-9},9.908\cdot10^{-9}]
\end{align*}
}
where (only quadratic terms shown for brevity)
\begingroup
\small
\setlength{\jot}{0pt}
\medmuskip=2mu
\begin{align*}
u_1(z,t)={}&1.4+0.15z_1+2.4t+0.05tz_2-1.852t^2-0.024t^2z_2-0.579t^2z_1\\
&{}-0.0105t^2z_1z_2-0.027t^2z_1^2-0.0005625t^2z_1^2z_2+O(t^3)\\[3pt]
u_2(z,t)={}&2.4+0.05z_2-3.704t-0.048tz_2-1.158tz_1-0.021tz_1z_2-0.054tz_1^2\\
&{}-0.001125tz_1^2z_2-7.486t^2-0.338t^2z_2-0.0035t^2z_2^2+0.4697t^2z_1\\
&{}-0.01584t^2z_1z_2-0.000375t^2z_1z_2^2+0.3108t^2z_1^2+0.00549t^2z_1^2z_2\\
&{}+0.02437t^2z_1^3+0.0004725t^2z_1^3z_2+0.0006075t^2z_1^4+0.00001266t^2z_1^4z_2+O(t^3)
\end{align*}
\endgroup
\end{example}

\section{Experimental Results}
\label{sec:experimental_results}
Two proof checkers were implemented to formally verify certificates of Picard-valid Taylor model traces (Appendix \ref{app:implementation_details} for more details), along with accompanying proof extractors for the tools Ariadne \footnote{Specifically the $\texttt{TaylorPicardIntegrator}$ algorithm which uses Taylor models.} \cite{DBLP:conf/adhs/CollinsBGV12}, Flow* \cite{DBLP:conf/cav/ChenAS13} and JuliaReach \cite{DBLP:conf/hybrid/BogomolovFFPS19}.
The first proof checker is a tactic-based implementation in \KeYmaeraX \cite{Fulton_Mitsch_Quesel_Volp_Platzer_2015}, a general-purpose theorem prover for \dL. In particular, this implementation follows the completeness result in Section\,\ref{sec:proof_checking} faithfully, and does not require modifying the underlying soundness kernel, enabling the reliable certification of numerical approximations using \KeYmaeraX's small trusted kernel \cite{Fulton_Mitsch_Quesel_Volp_Platzer_2015}. The second is an implementation in Lean4 \cite{DBLP:conf/cade/Moura021} based on reflection\footnote{Motivated by recent work in certifying LRAT proofs with reflection in Lean4 \cite{DBLP:journals/corr/lrat_catcher}.}, building upon a deep embedding of \dL in Lean4 \cite{dLean}. The goal of this checker is to provide efficient certification of Taylor model traces via a (provably correct) compiled program, such that benchmark instances from the ARCH-COMP \cite{ARCH26:ARCH_COMP26_Category_Report} can be provably certified. As this checker runs a compiled program, it relies on \verb|native_decide| to check Taylor model certificates without explicitly reconstructing the proof terms. Partly due to this, the implementation in Lean4 is much more efficient and scalable.

Finally, the Picard certificates (Definition \ref{def:picard_certificate}) generated by external tools are generally quite large, and it is therefore difficult to audit the faithfulness of the formally certified proof. To this end, the proof checkers further implement the functionality of formally verifying user-supplied safety conditions. That is, given a desired safety condition $S(x)$ for a reachability problem $\cR$, the proof checkers can formally verify the \dL formula 
\[x \in \X \land t = 0 \rightarrow \dbox{x' = p(x), t' = 1 \& t \leq T}{S(x)}\]
Internally, the proof checkers first verify the Picard-certificate of the Taylor model trace that was computed externally, and then check that the desired safety condition is implied by the Taylor model trace. In essence formally verifying the soundness of the following \dL formulas 

\[
    \vdash_{\dL} \fR \in \T = (\T_0, \dots, \T_N) \qquad \vdash_{\dL} \big(\bigvee_{i} x \in \T_i(B_i, [0, t_{i + 1} - t_i]) \big) \rightarrow S(x)
\]

\subsection{Benchmark}
The reachability problems are taken from the benchmarks used in ARCH COMP \cite{ARCH26:ARCH_COMP26_Category_Report} within the past 10 years in the nonlinear dynamics track with polynomial vector fields, resulting in a total of 10 safety problems (Appendix \ref{app:additional_experimental_results}, Table \ref{tab:benchmark-goals}), out of which 8 could be solved by at least one tool. The four ODE families are Van der Pol (VDP), coupled Van der Pol (CVDP), Laub--Loomis (LL), and Robertson (ROB), the exact parameter specifications are listed in Table \ref{tab:benchmark-goals}:
{\small
\[
\begin{aligned}
\mathrm{VDP}_{\mu}:\quad
  &x'=y \qquad y'=\mu(1-x^2)y-x \\[3pt]
\mathrm{CVDP}_{\mu}:\quad
  &x_i'=y_i \qquad y_i'=\mu(1-x_i^2)y_i-2x_i+x_{3-i}
    \qquad (i\in\{1,2\}) \\[3pt]
\mathrm{LL}_{W}:\quad
  &\begin{aligned}
    x_1'&=1.4x_3-0.9x_1 &\quad x_2'&=2.5x_5-1.5x_2 \\
    x_3'&=0.6x_7-0.8x_2x_3 & x_4'&=2-1.3x_3x_4 \\
    x_5'&=0.7x_1-x_4x_5 & x_6'&=0.3x_1-3.1x_6 \\
    x_7'&=1.8x_6-1.5x_2x_7
  \end{aligned} \\[3pt]
\mathrm{ROB}_{k}:\quad
  &x'=-0.4x+\beta_k yz \qquad
    y'=0.4x-\beta_k yz-\gamma_k y^2 \qquad z'=\gamma_k y^2
\end{aligned}
\]
}

\subsection{Results}

\newcommand{\oright}{\textcolor{vgreen}{$\bm{\checkmark}$}}%
\newcommand{\owrong}{\textcolor{vred}{$\bm{\lightning}$}}%
\begin{table}[tbh]
\centering
\caption{Lean proof-checking results for the 17 exported Taylor model traces:
8 proved (\oright) and 9 rejected (\owrong).}
\label{tab:lean-results}
\small
\setlength{\tabcolsep}{3pt}
\renewcommand{\arraystretch}{1.2}
\begin{tabular}{|l|l|r|r|r|c|}
\hline
\textbf{Instance} & \textbf{Tool} & \textbf{Steps}
& \shortstack{\textbf{Size}\\(MiB)}
& \shortstack{\textbf{Time}\\(s)}
& \textbf{Result} \\
\hline
$\bm{\mathrm{VDP}_{1}}$
& \textbf{Ariadne} & $\mathbf{372}$ & \textbf{5.73} & \textbf{15.74} & \oright \\
\cline{2-6}
& Flow* & $175$ & 1.24 & 7.55 & \owrong \\
\hline
$\bm{\mathrm{CVDP}_{1}}$
& \textbf{Ariadne} & $\mathbf{1\,427}$ & \textbf{229.07} & \textbf{398.95} & \oright \\
\cline{2-6}
& Flow* & $350$ & 5.85 & 8.72 & \owrong \\
\cline{2-6}
& JuliaReach & $211$ & 4.53 & 9.59 & \owrong \\
\hline
$\bm{\mathrm{LL}_{0.01}}$
& \textbf{Ariadne} & $\mathbf{217}$ & \textbf{16.15} & \textbf{31.39} & \oright \\
\cline{2-6}
& Flow* & $200$ & 4.30 & 8.10 & \owrong \\
\cline{2-6}
& JuliaReach & $266$ & 15.15 & 11.00 & \owrong \\
\hline
$\bm{\mathrm{LL}_{0.05}}$
& \textbf{Ariadne} & $\mathbf{375}$ & \textbf{76.60} & \textbf{137.90} & \oright \\
\cline{2-6}
& JuliaReach & $271$ & 8.30 & 8.88 & \owrong \\
\hline
$\bm{\mathrm{LL}_{0.1}}$
& \textbf{Ariadne} & $\mathbf{329}$ & \textbf{163.33} & \textbf{586.42} & \oright \\
\cline{2-6}
& Flow* & $400$ & 12.08 & 8.93 & \owrong \\
\cline{2-6}
& JuliaReach & $349$ & 10.69 & 9.16 & \owrong \\
\hline
$\bm{\mathrm{ROB}_{1}}$
& \textbf{Ariadne} & $\mathbf{10\,001}$ & \textbf{61.05} & \textbf{103.32} & \oright \\
\cline{2-6}
& JuliaReach & $33\,431$ & 110.65 & 50.63 & \owrong \\
\hline
$\bm{\mathrm{ROB}_{2}}$
& \textbf{Ariadne} & $\mathbf{66\,147}$ & \textbf{421.13} & \textbf{758.59} & \oright \\
\hline
$\bm{\mathrm{ROB}_{3}}$
& \textbf{Ariadne} & $\mathbf{134\,888}$ & \textbf{766.55} & \textbf{1833.35} & \oright \\
\hline
\multicolumn{6}{|p{\dimexpr\linewidth-2\tabcolsep-2\arrayrulewidth\relax}|}{%
\footnotesize
Lean4 checker, locally run on a machine with an AMD 5600 CPU with 32GB RAM.} \\
\hline
\end{tabular}
\end{table}

\begin{figure}[!tbh]
  \centering
  \includegraphics[width=0.8\linewidth]{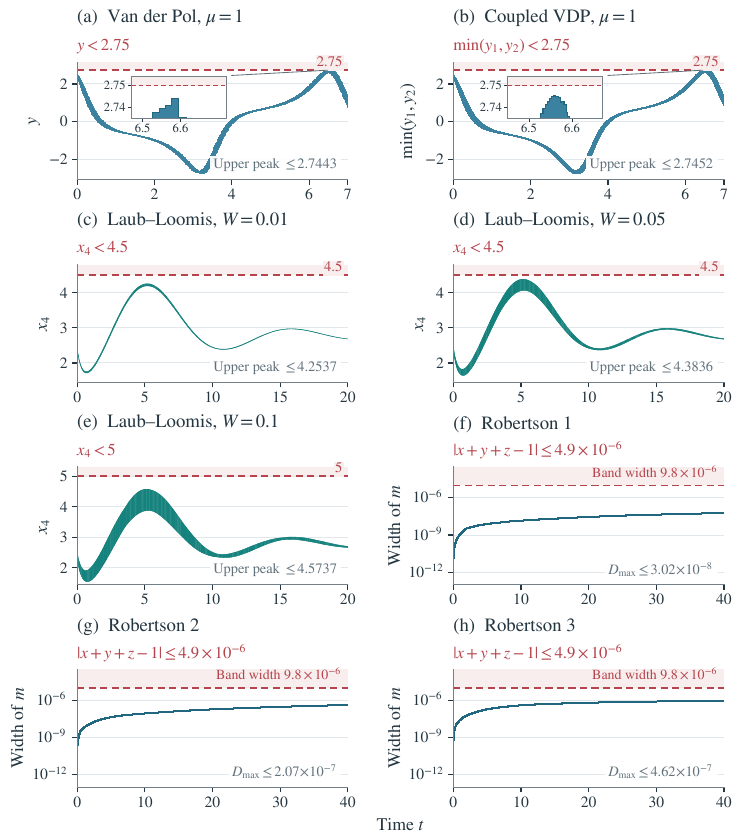}
  \caption{Eight Ariadne runs certified by Lean4, with safety conditions
    shown in red. Panels (f)--(h) plot enclosure widths of $m=x+y+z$,
    including all remainders; $D_{\max}$ bounds the deviation from $1$.}
  \label{fig:certified-safety}
\end{figure}

The proof extractors were able to extract a total of 17 Picard-certificates across the tools (Appendix \ref{app:additional_experimental_results}), the Lean4 proof checker was able to certify 8 of the 17 extracted certificates, while rejecting the remaining 9. The results are summarized in Table \ref{tab:lean-results}. Across the problems, the external tools were run with the same configurations as used in ARCH COMP \cite{ARCH26:ARCH_COMP26_Category_Report} wherever available. The tool Ariadne was also run with custom settings (Table \ref{tab:solver-configurations}) on 5 instances attempting to produce valid enclosures certifying the desired safety conditions when ARCH COMP configurations were not available. 

Surprisingly, it was found that all Picard-certificates extracted from Flow* and JuliaReach could not be certified by the Lean4 proof checker. These certificates were \emph{unsound}, in the sense that the Picard-validity conditions in Theorem \ref{thm:picard_valid_trace} were not satisfied. For all but one of the unsound certificates, the very first step of the Taylor model trace was not Picard-valid (not contractive), and the Lean4 proof checker correctly rejected the certificates. This does not directly imply that the enclosures themselves are unsound, in theory they might still be valid enclosures without being Picard-valid (although we are unaware of such soundness arguments). Nonetheless, these findings revealed relevant implementation issues that could be utilized to demonstrate concrete unsound enclosures on other problems (Appendix \ref{app:additional_experimental_results}). On the other hand, Ariadne was impressively able to produce valid Picard certificates for all 8 problems that it was run on and was able to solve without proof extraction, and the Lean4 proof checker was able to certify all of them (Fig.\ref{fig:certified-safety}).

It is worth emphasizing that the takeaway from the experimental results is not to criticize the tools themselves, but rather to highlight the importance of formal verification of continuous reachability. The findings demonstrate that, due to the subtleties of differential equations, even state-of-the-art tools can produce unsound results, in part because of the use of intricate floating point algorithms, and therefore a proof checking approach is crucial for certifying the correctness of such approximations, both to aid the development of the tools themselves as well as to increase trust in their answers for safety-critical applications.

\begin{table}[t]
\centering
\begin{tabular}{|l|r|r|}
\hline
Instance & Verified / total steps & Time (s) \\
\hline
Robertson 1 & $10 / 10\,001$  & 458.7 \\
Robertson 2 & $10 / 66\,147$  & 527.6 \\
Robertson 3 & $10 / 134\,888$ & 516.7 \\
\hline
\end{tabular}
\caption{KeYmaera X verification of the first 10 steps of the Ariadne
certificates for Robertson problems.}
\label{tab:kyx-robertson}
\end{table}
In comparison to the Lean4 implementation, the \KeYmaeraX implementation was much less efficient, unable to complete even a single step of the Taylor model trace within a reasonable time frame for most cases, except for the Robertson benchmarks (Table \ref{tab:kyx-robertson}). It appears that this difference in efficiency is primarily due to the lack of support for fast real arithmetic. \KeYmaeraX was not implemented with a focus on concrete number real arithmetic, and relies on its underlying quantifier-elimination procedure (Z3/Mathematica) to discharge real arithmetic goals. This is much less efficient than the Lean4 implementation that compiles a dedicated proof checker for such Taylor model traces using Lean4's (formally verified) floating point arithmetic \cite{DBLP:journals/corr/abs-2609-19352}.

In the future, it would be interesting to investigate optimizations to the \KeYmaeraX implementation, such that it is also capable of efficiently certifying Taylor model traces. 
Exporting its resulting proof terms would then also make it possible to verify the symbolic proofs in other \dL formalizations such as in Isabelle/HOL and Rocq \cite{DBLP:conf/cpp/BohrerRVVP17}, and PVS \cite{dl_pvs}.

\section{Related Works}
To the best of our knowledge, this is the first work for proof checking of continuous reachability. Relevant works concerning the formal verification of continuous reachability and the deductive proof theory of ODEs are detailed below. 

\noindent \emph{Formal Verification of Continuous Reachability:} The rigorous numerical validation of overapproximations for ODEs is a well-studied problem, resulting in many efficient implementations of tools \cite{DBLP:conf/cpsweek/Althoff15,DBLP:conf/cav/ChenAS13,DBLP:conf/hybrid/BogomolovFFPS19,Sandretto_2016_VEI,DBLP:conf/adhs/CollinsBGV12,DBLP:journals/jar/Immler18} utilizing algorithms such as set-based abstractions \cite{DBLP:conf/cpsweek/Althoff15}, rigorous Runge-Kutta methods \cite{DBLP:journals/jar/Immler18,Sandretto_2016_VEI} and Taylor models \cite{DBLP:conf/adhs/CollinsBGV12,DBLP:conf/cav/ChenAS13,DBLP:conf/hybrid/BogomolovFFPS19}. The abundance of tools also resulted in an annual competition \cite{ARCH26:ARCH_COMP26_Category_Report} that evaluates their performances across a variety of benchmarks. Outside a few exceptions \cite{DBLP:journals/jar/Immler18,DBLP:conf/cpp/Thies26}, such algorithms are generally not formally verified. In particular, existing approaches formally verify particular algorithms' implementations, whereas the proof checking approach presented in this article formally verifies the \emph{outputs} of approximations, thereby entirely decoupling the correctness certification from the evolving implementation of the algorithms themselves. 
\\
\noindent \emph{Deductive Verification of ODEs:} The deductive verification of ODEs aims to derive desired properties of ODEs from a small set of sound axioms \cite{DBLP:journals/jar/Platzer08,DBLP:journals/jar/Platzer17}, thereby obtaining trustworthy proofs of correctness that can be mechanically verified. Earlier works in this area have established fundamental properties of \dL's axiomatization, such as soundness \cite{DBLP:journals/jar/Platzer08}, relative completeness \cite{DBLP:journals/jar/Platzer08,DBLP:conf/lics/Platzer12a,DBLP:journals/jar/Platzer17}, completeness for ODE invariants \cite{DBLP:journals/jacm/PlatzerT20}, and completeness for compact IVPs \cite{DBLP:journals/jacm/PlatzerQ25}.
Furthermore, \dL has been successfully applied to formally verify delicate dynamical properties, such as stability \cite{DBLP:conf/hybrid/TanMP22} and liveness \cite{DBLP:journals/fac/TanP21}. More recently, it has been established that \dL is complete for the deductive verification of reachability problems \cite{DBLP:journals/jacm/PlatzerQ25}.
An interesting observation from Section\,\ref{sec:proof_checking} is that \dL is complete for contractive Taylor models, which generally relies on Schauder's fixed point theorem. 

\section{Conclusion and Future Works}

This paper presented a proof checking approach to certifying approximations for continuous reachability. Establishing the completeness of \dL for Taylor model enclosures, as well as prototype implementations of proof checkers in \KeYmaeraX and Lean4. The latter implementation is demonstrated to be capable of formally certifying proofs of safety for problems used in ARCH COMP \cite{ARCH26:ARCH_COMP26_Category_Report}. The findings of unsound results in Section\,\ref{sec:experimental_results} also highlight the intricacy of such algorithms. 

Just like proof checking was highly impactful for helping SAT solvers to be both efficient and correct, this work has the potential to revolutionize the reliability of continuous reachability tools.
For future works, it would be interesting to extend the proof checkers to handle all state-of-the-art tools participating in ARCH COMP, additionally including set-based abstractions \cite{DBLP:conf/cpsweek/Althoff15} and Runge-Kutta methods \cite{Sandretto_2016_VEI}.

\section*{AI Usage Disclosure}
AI tools (GPT Astra) were used in the implementations of the proof checkers, generation of figures, and proofreading the manuscript. The completeness theorem in Section\,\ref{sec:proof_checking} and architecture of the proofs were developed by the authors. 

\bibliographystyle{splncs04}
\bibliography{references}

\begin{thebibliography}{10}
\providecommand{\url}[1]{\texttt{#1}}
\providecommand{\urlprefix}{URL }
\providecommand{\doi}[1]{https://doi.org/#1}

\bibitem{DBLP:conf/cpsweek/Althoff15}
Althoff, M.: An introduction to {CORA} 2015. In: Frehse, G., Althoff, M. (eds.) 1st and 2nd International Workshop on Applied veRification for Continuous and Hybrid Systems, ARCH@CPSWeek 2014, Berlin, Germany, April 14, 2014 / ARCH@CPSWeek 2015, Seattle, WA, USA, April 13, 2015. EPiC Series in Computing, vol.~34, pp. 120--151. EasyChair (2015). \doi{10.29007/ZBKV}

\bibitem{Alur_2015}
Alur, R.: Principles of Cyber-Physical Systems. The MIT Press (2015)

\bibitem{BiereFallerFazekasFleuryFroleyksPollitt-SAT-Competition-2024-solvers}
Biere, A., Faller, T., Fazekas, K., Fleury, M., Froleyks, N., Pollitt, F.: {CaDiCaL}, {Gimsatul}, {IsaSAT} and {Kissat} entering the {SAT Competition 2024}. In: Heule, M., Iser, M., J{\"a}rvisalo, M., Suda, M. (eds.) Proc.~of {SAT Competition} 2024 -- Solver, Benchmark and Proof Checker Descriptions. Department of Computer Science Report Series B, vol. B-2024-1, pp. 8--10. University of Helsinki (2024)

\bibitem{DBLP:conf/hybrid/BogomolovFFPS19}
Bogomolov, S., Forets, M., Frehse, G., Potomkin, K., Schilling, C.: {JuliaReach}: a toolbox for set-based reachability. In: Ozay, N., Prabhakar, P. (eds.) Proceedings of the 22nd {ACM} International Conference on Hybrid Systems: Computation and Control, {HSCC} 2019, Montreal, QC, Canada, April 16-18, 2019. pp. 39--44. {ACM} (2019). \doi{10.1145/3302504.3311804}

\bibitem{DBLP:conf/cpp/BohrerRVVP17}
Bohrer, B., Rahli, V., Vukotic, I., V{\"o}lp, M., Platzer, A.: Formally verified differential dynamic logic. In: Bertot, Y., Vafeiadis, V. (eds.) Certified Programs and Proofs - 6th ACM SIGPLAN Conference, CPP 2017, Paris, France, January 16-17, 2017. pp. 208--221. {ACM} (2017). \doi{10.1145/3018610.3018616}

\bibitem{DBLP:conf/cav/ChenAS13}
Chen, X., {\'{A}}brah{\'{a}}m, E., Sankaranarayanan, S.: Flow*: An analyzer for non-linear hybrid systems. In: Sharygina, N., Veith, H. (eds.) Computer Aided Verification - 25th International Conference, {CAV} 2013, Saint Petersburg, Russia, July 13-19, 2013. Proceedings. LNCS, vol.~8044, pp. 258--263. Springer (2013). \doi{10.1007/978-3-642-39799-8\_18}

\bibitem{DBLP:conf/adhs/CollinsBGV12}
Collins, P., Bresolin, D., Geretti, L., Villa, T.: Computing the evolution of hybrid systems using rigorous function calculus. In: Heemels, M., Schutter, B.D. (eds.) 4th {IFAC} Conference on Analysis and Design of Hybrid Systems, {ADHS} 2012, Eindhoven, The Netherlands, June 6-8, 2012. {IFAC} Proceedings Volumes, vol.~45, pp. 284--290. Elsevier (2012). \doi{10.3182/20120606-3-NL-3011.00063}

\bibitem{Sandretto_2016_VEI}
{dit Sandretto}, J.A., Chapoutot, A.: Validated explicit and implicit {R}unge--{K}utta methods. Reliable Computing  \textbf{22}(1),  79--103 (Jul 2016)

\bibitem{dLean}
Enes, O., Prebet, E.: {dLean}: Uniform substitution for differential dynamic logic in {Lean}. Software repository (2026), \url{https://github.com/eneoli/dLean}, revision \texttt{902f778}

\bibitem{Fulton_Mitsch_Quesel_Volp_Platzer_2015}
Fulton, N., Mitsch, S., Quesel, J.D., Völp, M., Platzer, A.: {KeYmaera X}: An Axiomatic Tactical Theorem Prover for Hybrid Systems, LNCS, vol.~9195, p. 527–538. Springer International Publishing, Cham (2015). \doi{10.1007/978-3-319-21401-6_36}

\bibitem{DBLP:journals/corr/abs-2609-19352}
George, R.J., Adkisson, W., Anandkumar, A.: Floatlib: Verified floating-point arithmetic in lean. CoRR  \textbf{abs/2609.19352} (2026). \doi{10.48550/ARXIV.2609.19352}

\bibitem{ARCH26:ARCH_COMP26_Category_Report}
Geretti, L., Sandretto, J.A.D., Althoff, M., Benet, L., Collins, P., Forets, M., Mitsch, S., Perschl, M., Schilling, C., Tillet, J., Wooding, B.: Arch-comp26 category report: Continuous and hybrid systems with nonlinear dynamics. In: Frehse, G., Althoff, M., Geretti, L. (eds.) Proceedings of 13th Int. Workshop on Applied Verification for Continuous and Hybrid Systems. EPiC Series in Computing, vol.~110, pp. 52--84. EasyChair (2026). \doi{10.29007/f5l7}

\bibitem{DBLP:journals/jar/Immler18}
Immler, F.: A verified {ODE} solver and the {Lorenz} attractor. J. Autom. Reason.  \textbf{61}(1-4),  73--111 (2018). \doi{10.1007/S10817-017-9448-Y}

\bibitem{DBLP:conf/snc/MakinoB09}
Makino, K., Berz, M.: Rigorous integration of flows and {ODEs} using {Taylor} models. In: Kai, H., Sekigawa, H., Sasaki, T., Shirayanagi, K., Kotsireas, I.S. (eds.) Symbolic Numeric Computation, {SNC} '09, Kyoto, Japan - August 03 - 05, 2009. pp. 79--84. {ACM} (2009). \doi{10.1145/1577190.1577206}

\bibitem{Mitra21}
Mitra, S.: Verifying Cyber-Physical Systems: A Path to Safe Autonomy. MIT Press (2021)

\bibitem{DBLP:conf/cade/Moura021}
de~Moura, L., Ullrich, S.: The lean 4 theorem prover and programming language. In: Platzer, A., Sutcliffe, G. (eds.) Automated Deduction - {CADE} 28 - 28th International Conference on Automated Deduction, Virtual Event, July 12-15, 2021, Proceedings. Lecture Notes in Computer Science, vol. 12699, pp. 625--635. Springer (2021). \doi{10.1007/978-3-030-79876-5\_37}

\bibitem{tm_neher_jackson_nedialkov}
Neher, M., Jackson, K.R., Nedialkov, N.S.: On taylor model based integration of odes. SIAM Journal on Numerical Analysis  \textbf{45}(1),  236--262 (2007). \doi{10.1137/050638448}

\bibitem{DBLP:journals/jar/Platzer08}
Platzer, A.: Differential dynamic logic for hybrid systems. J. Autom. Reason.  \textbf{41}(2),  143--189 (2008). \doi{10.1007/S10817-008-9103-8}

\bibitem{DBLP:conf/lics/Platzer12a}
Platzer, A.: The complete proof theory of hybrid systems. In: Proceedings of the 27th Annual {IEEE} Symposium on Logic in Computer Science, {LICS} 2012, Dubrovnik, Croatia, June 25-28, 2012. pp. 541--550. {IEEE} Computer Society (2012). \doi{10.1109/LICS.2012.64}

\bibitem{DBLP:journals/jar/Platzer17}
Platzer, A.: A complete uniform substitution calculus for differential dynamic logic. J. Autom. Reason.  \textbf{59}(2),  219--265 (2017). \doi{10.1007/S10817-016-9385-1}

\bibitem{DBLP:journals/jacm/PlatzerQ25}
Platzer, A., Qian, L.: Axiomatization of compact initial value problems: Open properties. J. {ACM}  \textbf{72}(6),  1--51 (2025). \doi{10.1145/3763228}

\bibitem{DBLP:journals/jacm/PlatzerT20}
Platzer, A., Tan, Y.K.: Differential equation invariance axiomatization. J. {ACM}  \textbf{67}(1),  6:1--6:66 (2020). \doi{10.1145/3380825}

\bibitem{DBLP:conf/sat/PollittFFFSSB26}
Pollitt, F., Fleury, M., Fazekas, K., Froleyks, N., Schidler, A., Schreiber, D., Biere, A.: {CaDiCaL} 3.0 (tool paper). In: Ignatiev, A., Szeider, S. (eds.) 29th International Conference on Theory and Applications of Satisfiability Testing, {SAT} 2026, Lisbon, Portugal, July 20-23, 2026. LIPIcs, vol.~377, pp. 40:1--40:14. Schloss Dagstuhl - Leibniz-Zentrum f{\"{u}}r Informatik (2026). \doi{10.4230/LIPICS.SAT.2026.40}

\bibitem{dl_pvs}
Slagel, J.T., Mu\~{n}oz, C., Balachandran, S., Moscato, M., Dutle, A., Masci, P., White, L.: Towards an implementation of differential dynamic logic in pvs. In: Proceedings of the 11th ACM SIGPLAN International Workshop on the State Of the Art in Program Analysis. p. 44–50. SOAP 2022, Association for Computing Machinery, New York, NY, USA (2022). \doi{10.1145/3520313.3534661}

\bibitem{DBLP:journals/corr/lrat_catcher}
Szeider, S.: Lrat-catcher: Importing {SAT} solver certificates into lean4 by reflection. CoRR  \textbf{abs/2607.00815} (2026). \doi{10.48550/ARXIV.2607.00815}

\bibitem{DBLP:journals/sttt/TanHM23}
Tan, Y.K., Heule, M.J.H., Myreen, M.O.: Verified propagation redundancy and compositional {UNSAT} checking in {CakeML}. Int. J. Softw. Tools Technol. Transf.  \textbf{25}(2),  167--184 (2023). \doi{10.1007/S10009-022-00690-Y}

\bibitem{DBLP:conf/hybrid/TanMP22}
Tan, Y.K., Mitsch, S., Platzer, A.: Verifying switched system stability with logic. In: Bartocci, E., Putot, S. (eds.) {HSCC} '22: 25th {ACM} International Conference on Hybrid Systems: Computation and Control, Milan, Italy, May 4 - 6, 2022. pp. 2:1--2:11. {ACM} (2022). \doi{10.1145/3501710.3519541}

\bibitem{DBLP:journals/fac/TanP21}
Tan, Y.K., Platzer, A.: An axiomatic approach to existence and liveness for differential equations. Formal Aspects Comput.  \textbf{33}(4-5),  461--518 (2021). \doi{10.1007/S00165-020-00525-0}

\bibitem{DBLP:conf/cpp/Thies26}
Thies, H.: Computing solutions for systems of multivariate ordinary differential equations in {Rocq}. In: Stark, K., Zakowski, Y., Swamy, N., Tabareau, N. (eds.) Proceedings of the 15th {ACM} {SIGPLAN} International Conference on Certified Programs and Proofs, {CPP} 2026, Rennes, France, January 12-13, 2026. pp. 29--44. {ACM} (2026). \doi{10.1145/3779031.3779097}

\bibitem{Walter_1998}
Walter, W.: Ordinary Differential Equations, Graduate Texts in Mathematics, vol.~182. Springer New York (1998). \doi{10.1007/978-1-4612-0601-9}

\bibitem{DBLP:conf/sat/WetzlerHH14}
Wetzler, N., Heule, M., Jr., W.A.H.: Drat-trim: Efficient checking and trimming using expressive clausal proofs. In: Sinz, C., Egly, U. (eds.) Theory and Applications of Satisfiability Testing - {SAT} 2014 - 17th International Conference, Held as Part of the Vienna Summer of Logic, {VSL} 2014, Vienna, Austria, July 14-17, 2014. Proceedings. Lecture Notes in Computer Science, vol.~8561, pp. 422--429. Springer (2014). \doi{10.1007/978-3-319-09284-3\_31}

\end{thebibliography}

\clearpage
\appendix
\section{Proofs}
\label{app:proofs}

\subsection{One-Step Completeness}
\begin{proof}[Theorem~\ref{thm:one_step_completeness}]
Following the proof sketch, it remains to show the existence of $\T'$.
On the compact domain $\X\times[0,T]$, the $\max/\min$ of
$p(r(z,t)+I)$ are continuous and can be approximated arbitrarily closely
by rational polynomials. Integrating the $\max/\min$ expressions gives the componentwise bounds of the Picard image. The positive margin $\eps \in \Q^+$ from Picard-validity therefore implies that sufficiently accurate polynomial approximations preserve strict containment, hence a desired $\T'$ necessarily exists, as desired. 
\qedhere
\end{proof}

\subsection{Preconditioned Completeness}
\begin{proof}[Theorem~\ref{thm:preconditioned_completeness}]
Recall that the desired goal is to prove
\[
    x\in\X\land t=0 \rightarrow
    \dbox{x'=p(x),\ t'=1\ \&\ t\leq T}{x\in\T(B, t)}
\]
The proof directly follows from the one-step case:
\begin{itemize}
    \item Apply the one-step argument with constant parameters $z,a$.
    Local Picard-validity establishes
    {
    \[
        z\in D\land x\in\cA(z) \land t=0\ \rightarrow 
            \dbox{x'=p(x),\ t'=1\ \&\ t\leq T}{x\in\cU(z,t)}
    \]
    }
    \item By $\X\subseteq\cA(\cV(B))$, choose witnesses $b\in B$ and
    $z\in\cV(b)$ for the initial state $x \in\cA(z)$. 
    \item The one-step argument gives $x\in\cU(\cV(B),t)$ along the flow, the previous step then gives the desired derivation. The preconditioning obligations are formulas
    $\folr$, hence provable with quantifier elimination. 
    \qedhere
\end{itemize}
\end{proof}

\subsection{Completeness for Traces}
To complete the proof of Theorem \ref{thm:completeness_of_tm}, the following proof rule is needed. 

\begin{theorem}[\cite{DBLP:journals/jacm/PlatzerQ25}]
    The following is a derivable proof rule of \dL. 

    \begin{calculus}
        \cinferenceRule[stepExt|StepExt]{extending solutions}
        {\medmuskip=3mu
         \thickmuskip=2.75mu
         \linferenceRule[tableaux]
            {\lsequent{t = t_0, P_1}{\Gamma_2}\\[0.3ex]
             \lsequent{\Gamma_1}{[x' = f(x), t' = 1\,\&\,t \leq t_0]P_1}\\[0.3ex]
             \lsequent{\Gamma_2}{[x' = f(x), t' = 1\,\&\,t \leq t_0 + t_1]P_2}}
            {\lsequent{t \leq t_0, \Gamma_1}{[x' = f(x), t' = 1\,\&\,t \leq t_0 + t_1]\bigl((t \leq t_0 \rightarrow P_1) \land (t > t_0 \rightarrow P_2)\bigr)}}
        }{}
    \end{calculus}
\end{theorem}

\begin{proof}[Theorem~\ref{thm:completeness_of_tm}, outline]
Let $h_i=t_{i+1}-t_i$, the desired goal is
\[  
    x\in\X\land t=0\ \rightarrow
        \dbox{x'=p(x),\ t'=1\ \&\ t\leq T}
        {\bigwedge_{i=0}^{N}\bigl(t_i\leq t\leq t_{i+1}\rightarrow x \in \T_i(B_i, t - t_i)\bigr)}
\]
The proof amounts to chaining up preconditioned completeness with the proof rule \irref{stepExt}:
\begin{itemize}
    \item Applying preconditioned completeness at the $i$-th step gives
    \[
        x\in E_i\land t=t_i\ \rightarrow
        \dbox{x'=p(x),\ t'=1\ \&\ t\leq t_{i+1}}{x \in \T_i(B_i, t - t_i)}
    \]
    \item The handoff condition
    $\T_{i-1}(B_{i-1},h_{i-1})\subseteq E_i$ is a formula of real arithmetic and therefore provable with quantifier elimination.
    \item Starting with $\X\subseteq E_0$, apply \irref{stepExt}
    iteratively to concatenate the safety conditions across the time steps $[0, t_1], \dots, [t_N, t_{N + 1}]$. Thus the resulting proof covers the overall time step $[0, T]$, as desired. 
    \qedhere
\end{itemize}
\end{proof}

\section{Additional Implementation Details}
\label{app:implementation_details}

\subsection{Lean4 Checker}

The proof checker implemented in Lean4 is primarily based on a concrete compiled program \verb|check| that provably satisfies the following Lean4 theorem 

\[\forall \pi \forall \Phi \in \text{SAFE}_{\dL}~\texttt{check}(\pi, \Phi) = \texttt{true} \rightarrow \texttt{sound}~\Phi\]

That is, if the algorithm returns \verb|true| on the safety formula $\Phi$ and untrusted certificate $\pi$, then it necessarily follows that $\Phi$ is a sound axiom of $\dL$ (thereby valid). In particular, this meta-level soundness theorem is proven once, and then \verb|checker| executed with \verb|native_decide| for efficient checking. 

Concretely, the checker leverages Lean4's floating point arithmetic with (formally verified) rigorous bounds \cite{DBLP:journals/corr/abs-2609-19352}, and directly checks the Picard-validity conditions (Theorem \ref{thm:picard_valid_trace}) on every preconditioned Taylor model, as well as the endpoint inclusions. The soundness theorem relating the execution of this program to the validity of $\Phi$ is then proven in Lean4 following the proofs of the completeness Theorems \ref{thm:one_step_completeness}, \ref{thm:preconditioned_completeness}, \ref{thm:completeness_of_tm}. For a formula $\Phi$ that is checked to be correct, \verb|#print axioms| outputs:
\begin{itemize}
    \item The standard \verb|propext|, \verb|Classical.choice|, \verb|Quot.sound|.
    \item One additional \verb|_native| axiom, indicating the usage of native evaluation. 
\end{itemize} 

The following is an example of the Van der Pol problem being checked. 

\begin{example}[Proof Checking VDP]
For the two-dimensional Van der Pol benchmark with $\mu=1$ (Table \ref{tab:benchmark-goals}), the desired safety condition is
{\small
\[5/4 \leq x \leq 31/20 \land 47/20 \leq y \leq 49/20 \land t=0 \rightarrow
\dbox{x'=y,\ y'=(1-x^2)y-x,\ t'=1\ \&\ t\leq7}{y<11/4}\]}
Let \verb|goal : Formula| be its encoding in dLean \cite{dLean}, and \verb|ariadne.tmseq| the Picard-certificate generated by Ariadne, then the validity/soundness of \verb|goal| can be checked with the implemented proof-checking command (via reflection), \verb|reachability_reflect_formula|, as follows:
\begingroup
\small
\begin{verbatim}
reachability_reflect_formula checked
  (goal) "ariadne.tmseq"

theorem safety : sound goal := checked

#check safety
#print axioms safety
\end{verbatim}
\endgroup
And if successful (which is indeed the case here), the safety theorem \verb|safety| is successfully established, proving the soundness/validity of \verb|goal|. The auditing commands also confirm that the only axioms used were the standard three plus an \verb|_native| axiom from the use of \verb|native_decide|. 
\end{example}

\subsection{\KeYmaeraX Checker}
The \KeYmaeraX checker was implemented to faithfully follow the completeness arguments in Section \ref{sec:proof_checking}. On a high level, this implementation consists of:

\begin{itemize}
    \item A tactic implemented in \KeYmaeraX, to automatically discharge Taylor model proof obligations. Note that this does not affect the soundness kernel. 
    \item An external implementation of an (untrusted) program $\mathcal{P}$, such that given a certificate $\pi$ and a safety formula $\Phi$, $\mathcal{P}(\pi, \Phi)$ is a \dL proof that can be checked by \KeYmaeraX. 
\end{itemize}

\section{Additional Experimental Results}

\label{app:additional_experimental_results}
\newcommand{\benchmath}[1]{%
  \raisebox{0pt}[\dimexpr\height+3pt\relax]
                 [\dimexpr\depth+3pt\relax]{$#1$}}

\begin{table}[htbp]
\centering
\caption{Ten benchmark instances: parameters, initial conditions at
$t=0$, and safety conditions on $[0,T]$.}
\label{tab:benchmark-goals}
\small
\setlength{\tabcolsep}{4pt}
\renewcommand{\arraystretch}{1.25}
\begin{tabular}{|l|c|c|c|c|}
\hline
\textbf{Family} & \textbf{Parameters} & \textbf{Initial condition}
& \textbf{Safety condition} & $T$ \\
\hline
$\mathrm{VDP}_{\mu}$
& $\mu=1$
& \benchmath{\begin{gathered}
  x\in[1.25,1.55]\\y\in[2.35,2.45]
  \end{gathered}}
& $y<2.75$ & $7$ \\
\cline{2-5}
& $\mu=2$
& \benchmath{\begin{gathered}
  x\in[1.55,1.85]\\y\in[2.35,2.45]
  \end{gathered}}
& $y<4$ & $8$ \\
\hline
$\mathrm{CVDP}_{\mu}$
& $\mu=1$
& \benchmath{\begin{gathered}
  x_1,x_2\in[1.25,1.55]\\y_1,y_2\in[2.35,2.45]
  \end{gathered}}
& $\min(y_1,y_2)<2.75$ & $7$ \\
\cline{2-5}
& $\mu=2$
& \benchmath{\begin{gathered}
  x_1,x_2\in[1.55,1.85]\\y_1,y_2\in[2.35,2.45]
  \end{gathered}}
& $\min(y_1,y_2)<4.05$ & $8$ \\
\hline
$\mathrm{LL}_{W}$
& $W\in\{0.01,0.05\}$
& $x\in c+[-W,W]^7$
& $x_4<4.5$ & $20$ \\
\cline{2-5}
& $W=0.1$
& $x\in c+[-W,W]^7$
& $x_4<5$ & $20$ \\
\hline
$\mathrm{ROB}_{k}$
& $k\in\{1,2,3\}$
& $(x,y,z)=(1,0,0)$
& \benchmath{\begin{gathered}
  |x+y+z-1|\\{}\leq 10^{-5}
  \end{gathered}}
& $40$ \\
\hline
\multicolumn{5}{|p{\dimexpr\linewidth-2\tabcolsep-2\arrayrulewidth\relax}|}{%
\footnotesize
For Laub--Loomis, $c=(1.2,1.05,1.5,2.4,1,0.1,0.45)$.
For Robertson, $(\beta_k,\gamma_k)=(10^2,10^3),(10^3,10^5),(10^3,10^7)$
for $k=1,2,3$.} \\
\hline
\end{tabular}
\end{table}

\begin{table}[htbp]
\centering
\caption{Solver configurations and candidate-extraction outcomes for the
ten benchmark instances. Years identify the reference ARCH settings.}
\label{tab:solver-configurations}
\small
\setlength{\tabcolsep}{4pt}
\renewcommand{\arraystretch}{1.2}
\begin{tabular}{|l|c|c|c|}
\hline
\textbf{Instance} & \textbf{Ariadne} & \textbf{Flow*}
& \textbf{JuliaReach} \\
\hline
$\mathrm{VDP}_{1}$ & Custom & Custom & --- \\
\hline
$\mathrm{VDP}_{2}$ & --- & --- & --- \\
\hline
$\mathrm{CVDP}_{1}$ & Custom & Custom & ARCH (2021) \\
\hline
$\mathrm{CVDP}_{2}$ & --- & --- & --- \\
\hline
$\mathrm{LL}_{0.01}$ & ARCH (2026) & Custom & ARCH (2026) \\
\hline
$\mathrm{LL}_{0.05}$ & ARCH (2026) & --- & ARCH (2026) \\
\hline
$\mathrm{LL}_{0.1}$ & ARCH (2026) & Custom & ARCH (2026) \\
\hline
$\mathrm{ROB}_{1}$ & Custom & --- & Custom \\
\hline
$\mathrm{ROB}_{2}$ & Custom & --- & --- \\
\hline
$\mathrm{ROB}_{3}$ & Custom & --- & --- \\
\hline
\multicolumn{4}{|p{\dimexpr\linewidth-2\tabcolsep-2\arrayrulewidth\relax}|}{%
\footnotesize
\textbf{ARCH}: certificate extracted using the ARCH COMP parameters.
\textbf{Custom}: certificate extracted with custom settings;
\textbf{---}: not run; no compatible setting was found to solve the safety problem.} \\
\hline
\end{tabular}
\end{table}

Table \ref{tab:solver-configurations} documents the extraction of Picard-certificates from the tools Ariadne, Flow* and JuliaReach, including the solver configurations used to generate the certificates, which are either the ARCH COMP settings or custom settings.

\subsection{Concrete Unsound Enclosures}
The following examples demonstrate unsound enclosures in JuliaReach and Flow*,     without proof extraction or source modifications.
Both use the scalar ODE
\[
    x'=x^2 \qquad x(0)=x_0 \qquad x(t)=\frac{x_0}{1-x_0t}
\]

\begin{example}[Flow* - commit
\href{https://github.com/chenxin415/flowstar/tree/b85a3211748cb77b736fe4ad42ee02d8d2b81148}{\texttt{b85a321}}]
Set the initial interval to $[7/8,9/8]$, the fixed step and requested
horizon to $1/1024$, and the Taylor-model order to $2$.
The native computation completes and bounds the entire returned step
above by
\[
    U=\frac{1268030346289239}{1125899906842624}
\]
However, take $x_0=9/8$ and
$t_*=1/1024-1/2^{30}$, which lies strictly inside the returned time domain.
The exact solution satisfies
\[
    x(t_*)=\frac{9663676416}{8580497417}
    \qquad
    x(t_*)-U
    =\frac{1265519231797959921}{9660781242463675857502208}>0
\]
The gap is approximately $1.31\cdot10^{-7}$. Moreover, supplying the
safety condition $x\leq9447563/8388608$ makes Flow* report \texttt{SAFE},
although the same trajectory violates it:
\[
    x(t_*)-\frac{9447563}{8388608}
    =\frac{3374224157}{71978429276225536}>0
\]
\end{example}

\begin{example}[JuliaReach]
With Julia~1.10.10 and TaylorModels.jl~0.10.2, we call
\texttt{validated\_integ2} (the validated integration solver used in JuliaReach) directly with $x(0)=1$ and $T=1/8$.
The settings are spatial order $2$, time order $3$,
\texttt{abstol=1e-3}, \texttt{parse\_eqs=false}, and $\delta=10^{-10}$,
with one spatial variable and jet order $4$.

Five steps complete without a validation warning, yielding the following enclosure at $T=1/8$
\[
    [1.1428553398650563,\ 1.1428553631885077]
\]
The exact solution $x(1/8)=8/7$ exceeds its upper endpoint by approximately
$1.78\cdot10^{-6}$:
\[
    U=\frac{643370373474267}{562949953421312}
    \qquad
    \frac87-U=\frac{7013050627}{3940649673949184}>0
\]
\end{example}

\end{document}